\documentclass[letterpaper, 10 pt, conference]{ieeeconf}  
\usepackage{FG2021}

\FGfinalcopy 

\IEEEoverridecommandlockouts                              
\usepackage{amssymb}
\usepackage{amsmath}
\usepackage{bm} 

\usepackage{graphicx}
\usepackage{hhline}
\usepackage{xcolor}
\usepackage{multirow}
\usepackage{wrapfig}

\def\FGPaperID{316} 

\title{\LARGE \bf
Predicting Trust Using Automated Assessment\\ of Multivariate Interactional Synchrony
}

\author{\parbox{16cm}{\centering
    {\large Adrien Meynard$^{1,2}$, Gayan Seneviratna$^3$, Elliot Doyle$^4$, Joyanne Becker$^3$, Hau-Tieng Wu$^{1,5}$, and Jana Schaich Borg$^3$}\\
    {\normalsize
    $^1$ Department of Mathematics, Duke University, Durham, USA\\
    $^2$ Univ Lyon, ENS de Lyon, CNRS, Laboratoire de Physique, Lyon, France\\
    $^3$ Social Science Research Institute, Duke University, Durham, USA\\
    $^4$ Department of Psychology, University of Oregon, Dayton, USA\\
    $^5$ Department of Statistical Science, Duke University, Durham, USA
    }}

    \thanks{This work was partially supported by Duke Bass Connections and Templeton World Charity Foundation grant TWCF0321.}
}

\newcommand{\xp}{\mathrm{xp}}
\newcommand{\AU}{\mathrm{AU}}

\def\bc{{\mathbf c}}
\def\bd{{\mathbf d}}
\def\bg{{\mathbf g}}

\def\bw{{\mathbf w}}
\def\bx{{\mathbf x}}
\def\by{{\mathbf y}}
\def\bz{{\mathbf z}}

\def\cD{{\mathcal D}}

\def\fs{f_{\mathrm{s}}}

\def\bbeta{\boldsymbol{\beta}}

\newcommand{\RR}{{\mathbb R}}

\begin{document}


\ifFGfinal
\thispagestyle{empty}
\pagestyle{empty}
\else
\author{Anonymous FG2021 submission\\ Paper ID \FGPaperID \\}
\pagestyle{plain}
\fi
\maketitle

\begin{abstract}

Diverse disciplines are interested in how the coordination of interacting agents’ movements, emotions, and physiology over time impacts social behavior.   Here, we describe a new multivariate procedure for automating the investigation of this kind of behaviorally-relevant ``interactional synchrony'', and introduce a novel interactional synchrony measure based on features of dynamic time warping (DTW) paths.  We demonstrate that our DTW path-based measure of interactional synchrony between facial action units of two people interacting freely in a natural social interaction can be used to predict how much trust they will display in a subsequent Trust Game.  We also show that our approach outperforms univariate head movement models, models that consider participants’ facial action units independently, and models that use previously proposed synchrony or similarity measures.  The insights of this work can be applied to any research question that aims to quantify the temporal coordination of multiple signals over time, but has immediate applications in psychology, medicine, and robotics.

\end{abstract}

\section{Introduction}
\label{se:Introduction}

When we interact with each other, our emotions, actions, movements, and physiology become coordinated over time~\cite{ vicaria2016meta}.  For example, when we walk down the street with another person we like, our steps fall into sync.  When we listen to somebody tell a story, we subconsciously imitate their facial expressions and gestures.  The interdependence between interacting people’s movements is referred to broadly as interactional synchrony~\cite{schirmer2021being} and has been shown to cause and reflect specific aspects of social communication and behavior, such as how much two people like each other or cooperate \cite{rennung2016prosocial}.  Knowing what types of interactional synchrony between what types of movements correlate with behaviors would provide insight into how human brains process social information, and could inform practical tools that screen for social disorders, predict negotiation outcomes, improve customer service interactions, or engender trust in social robots and avatars.  Towards that end, here, we propose a new automated method for measuring multivariate interactional synchrony in video interactions, and use it to identify facial features whose coordination predicts trust in the Trust Game, a game in which the earnings of two players are maximized when the first player trusts the second.

Automated interactional synchrony tools have proven difficult to develop \cite{delaherche2012interpersonal}.   One challenge is that social interaction data has traditionally been expensive and cumbersome to collect, resulting in small social synchrony data sets with only 30--40 interactions \cite{delaherche2012interpersonal}.  Another challenge is that many interactional synchrony researchers do not have access to automated image processing tools, so they extract and analyze features of an interaction by hand or through subjective labeling, which is extremely time-intensive.  Even when automated image analysis tools are available, the aspects of a social interaction that are most important to analyze are not clear.  Is synchrony between people's emotions, gestures, pose, or language most influential on the outcomes of social interactions?  Or do the interactions between all of these responses matter?  At present there are no well-established methods for assessing synchrony between multiple time series, so most interactional synchrony studies focus on a single feature, despite consensus from the psychology literature that social communication occurs through the simultaneous coordination of multiple types of non-verbal signals~\cite{vinciarelli2009social}.

Perhaps the greatest challenge to automating interactional synchrony measurements is that psychological definitions of interactional synchrony are not quantitative.  One definition preferred by the authors is “synchrony is the dynamic and reciprocal adaptation of the temporal structure of behaviors between interactive partners. Unlike mirroring or mimicry, synchrony is dynamic in the sense that the important element is the timing, rather than the nature of the behaviors”~\cite{delaherche2012interpersonal}.  Thus, interactional synchrony can incorporate mimicry or aligned oscillatory movements, but mimicry and oscillatory movements are neither necessary nor sufficient to explain the type of temporal coordination we know occurs between movements in real social interactions.  Another definition states ``synchrony refers to the motion interdependence of all participants during an interaction focusing on more than a single behavior''~\cite{kruzic2020facial}. This adds to the previous definition by clarifying that interactional synchrony must involve multiple interdependent motions, but leaves unspecified how the motions have to be interdependent.  It is not clear how to operationalize such vague concepts using traditional methods for assessing the coordination between two time series, because traditional methods often assume that only one type of temporal coordination (for example, only shared sinusoidal oscillations or only the time lags between two time series) is important.  Many traditional methods also assume the relationships between time series are stable over time.  The temporal coordination of real human social interactions, in contrast, is dynamic, changes directionality and cadence, and occurs differently for different types of movements (Person A may slowly follow Person B’s head pose while Person B simultaneously smiles almost immediately after they see Person A smile).  

The most popular method for automating social synchrony detection applies a two-stage analysis to videos of interactions.  First, motion energy analysis (MEA) is used to create a single time series of the frame-to-frame intensity differences of pixels from the head region of each participant.  Then social synchrony is assessed by analyzing the correlation coefficients 
of the windowed cross-correlations (WCC) of these paired time courses (by selecting the peak value or thresholding the coefficients to compute durations/bouts of synchrony).  A recent study demonstrated that the degree to which patients trusted clinicians in simulated interactions was mediated by social synchrony measured in this way \cite{goldstein2020clinician}, supporting the notion that at least some of the non-verbal cues that foster trust may be dynamically interdependent between actors and located somewhere in the head.  Yet, the univariate MEA method of measuring social synchrony has two important weaknesses.  First, it is unable to identify what facial expressions or specific head movements are coordinating in a behaviorally-relevant way.  Even if more detailed movements than overall head motion were extracted from the videos, procedures for multivariate interactional synchrony are not established.  Further, many of the multivariate methods one might consider assume the temporal relationships between all variables under investigation are the same, which we know is not true in social interactions, as discussed above.   Second, WCC assumes that the relationships between time series are stationary across the length of the chosen window.  When this assumption is unknowingly violated, the resulting correlation can be misleading \cite{boker2002windowed}.  As a consequence, WCC is not ideally-suited for highly dynamic temporal coordination, especially in multivariate settings where different features may have different optimal windows. 

Here, we propose a new procedure for automating the investigation of behaviorally-relevant social synchrony.  Rather than focus on single features of visual scenes, like MEA methods, our method accommodates multiple features of a social interaction and identifies which ones are behaviorally-relevant, even when the features are not fully independent.  We also introduce a new approach to measuring social synchrony that allows for dynamic time lags between actors and avoids the stationarity assumptions of WCC. We use dynamic time warping (DTW) to achieve this, but leverage characteristics of the warping path rather than the DTW ``distance''.  We demonstrate that our DTW-path based measure of social synchrony between facial action units of two people interacting freely in a natural social interaction can be used to predict how much they will trust each other in a subsequent Trust Game.  We also show that our approach outperforms univariate head movement synchrony models, models that only consider participants’ facial action units independently, and models that use WCC to assess synchrony.

\section{Data Collection}
\label{se:data}

\subsection{Overall Description}
Videos were recorded of two people in separate geographic locations interacting via Skype.  Each pair was given approximately three minutes to interact freely in any way they wished.  Although research assistants set up the session, they left the room during this free-interaction phase.  The research assistants returned to the room to open the Trust Game virtual interface (described below), and then left the room again so that the participants were alone when they played the Trust Game and filled out questionnaires.  135 pairs of videos were collected.  Our goal was to predict the outcome of the Trust Game using our assessment of each pair's interactional synchrony during the free interaction period.

\subsection{Trust Game}
In the Trust Game \cite{berg1995trust}, “H” (Head player) is given a dollar and given the opportunity to give \$0, \$0.20, \$0.40, \$0.60, \$0.80, or \$1 of that dollar to “T” (Tail player).  H is told that the amount they choose to give to T will be tripled before it is delivered.  After H makes their choice and the tripled amount is delivered, T is then given the opportunity to return to H as much as they want of the money they received.  H and T are aware of how much the other started with, and how much they choose to give.  The outcome of the game depends on H’s trust and T’s trustworthiness: to maximize earnings for both players, H would give T \$1 and trust that T would return more than \$1 of their earnings, and T would be trustworthy and follow-through with returning more than \$1 of their earnings.  H and T roles are randomly assigned.

\subsection{Facial Action Units}
Humans innately assess others’ trustworthiness when they see them, and use signals from the way others’ emotional expressions unfold over time to make these judgments \cite{todorov2009evaluating}.  Dynamic facial features play a more dominant role in trustworthiness judgments than static facial features \cite{ bonnefon2017can} or non-facial nonverbal cues like gestures or body posture \cite{kruzic2020facial}.  Therefore, we focused our analysis on the facial features that comprise players’ dynamic emotional expressions.  One option would be to track prototypic emotion categories, like “disgust” or “joy”.  However, the facial configurations associated with these emotion categories are not as consistent or universal as previously believed, calling into doubt currently available models for automatically tracking those types of supraordinate emotions  \cite{barrett2019emotional}.  Thus, we analyzed facial action units (AUs) instead of emotional categories, per se.  AUs are well-validated “minimal units of facial activity that are anatomically separate and visually distinguishable” \cite{ekman1997face}, such as lid raises or nose wrinkles.  Emotional expressions are comprised of multiple AUs working in tandem to different degrees in different people.  Automatic AU-detection is thought to be both more ecologically valid and more reliable than automatic emotion detection. We extracted the intensity (from zero to five) of the 17 AUs listed in Supplementary Table~S-I in each frame of each video in a pair of interactions using the open-source deep-neural-network (DNN) OpenFace \cite{baltrusaitis2018openface}.  OpenFace's confidence measures associated with each of its classifications were used in pre-processing.

\section{Procedure}
\label{se:Procedure}

\subsection{Notations}

$K_\AU$ denotes the number of action units (here, $K_\AU=17$). Index $k \in\{1,\ldots,K_\AU\}$ denotes the $k$th action unit from the list in Supplementary Table~S-I.  $N_\xp$ denotes the number of sessions. Index $n\in\{1,\ldots,N_\xp\}$ denotes the $n$th session. $M_n$ denotes the number of frames contained in the pair of video recordings of the natural interaction stage of session $n$. For a given session, $i\in\{1,2\}$ arbitrarily identifies the two subjects in a considered pair. The signal measuring the $k$th action unit of the $i$th subject of the $n$th session is therefore denoted by $\bx_{k,n}^{(i)}\in\RR^{M_n}$. Its $m$th sample is denoted by $\bx_{k,n}^{(i)}[m]$. The whole AU data set is thus comprised of $2K_\AU \sum_{n=1}^{N_\xp} M_n$ samples. As detailed in section~\ref{sse:binarization}, we try to predict a binarization of the H's choice in the Trust Game. This binary variable is denoted by $\by[n]$.

\subsection{Session Exclusions}

OpenFace provides a confidence score  $\bc_n^{(i)}\in\RR^{M_n}$ from $0$ to $1$ for its AU classifications in each frame. A session's quality is assessed via the worst confidence score over time:
\[
\bc_n^{(\min)}[m] = \min\left( \bc_n^{(1)}[m],\bc_n^{(2)}[m] \right),\quad\forall m\in\{1,\ldots,M_n\}.
\]
To reduce the impact of poor AU feature detection on the interactional synchrony assessments, we excluded sessions where the worst confidence score $\bc_n^{(\min)}$ was below a threshold $\tau$ for more than 30\% of the frames in at least one of the videos of a pair.  We set $\tau=0.7$, which resulted in twelve sessions being excluded. The rest of the analyses were performed on the remaining 123 sessions. 

\subsection{Step 1: Video Preprocessing}

\subsubsection{Smoothing}
\label{sse:smoothing}

The OpenFace model occasionally  detects facial landmarks or AUs inaccurately, particularly when a participant turns their head quickly or puts their hand in front of their face.  Such inaccuracies are brief, and can therefore cause artificial fast variations in the AU time series.  To prevent these artifacts from disproportionately influencing subsequent steps, we smoothed the AU time series.  $\tilde \bx_{k,n}^{(i)}$ denotes the smoothed version of the AU signal $\bx_{k,n}^{(i)}$. The smoothing is obtained via the moving average of the signal
\begin{equation}
\tilde \bx_{k,n}^{(i)}[m] = \dfrac1{|V_m|}\sum_{p\in V_m} \bx_{k,n}^{(i)}[p] ,
\end{equation}
where $V_m=\{m-\bd_n^{(i)}[m],\ldots,m+\bd_n^{(i)}[m]\}\cap\{1,\ldots,M_n\}$. Here, $\bd_n^{(i)}[m]$ denotes the smoothing half-width. This quantity is adjusted according to the OpenFace confidence score $\bc_n^{(i)}$. Thus, the smaller $\bc_n^{(i)}[m]$ is, the larger $\bd_n^{(i)}[m]$ is chosen. In practice, the dependence is linear:
\begin{equation}
\bd_{n}^{(i)}[m] = \left\lfloor d_{\max} - \left(d_{\max}-1\right)\bc_n^{(i)}[m] \right\rfloor,
\end{equation}
where $\lfloor\cdot\rfloor$ denotes the floor function, and $d_{\max}$ is the maximal permitted smoothing half-width. $d_{\max}$ is automatically chosen within the set $\{1,\ldots,\lfloor M_n/10 \rfloor\}$ as the optimal half-width which, when applied to the corresponding smoothing to the confidence score itself, gives the smoothed confidence score with the most frames above the threshold $\tau$.

\subsubsection{Optional Imputation of Low-Confidence Frames}
\label{sse:imputation}
We assessed the result of an optional preprocessing step that imputed AU values in frames where OpenFace's confidence estimate was less than the chosen value of $\tau$.  If imputation provides AU values that are more representative of ground truth than OpenFace's output of low confidence for those frames, interactional synchrony assessments may be improved.  We tested a linear imputation method. Assume that $\tilde \bx_{k,n}^{(i)}$ has to be imputed from sample $m_1$ to sample $m_2$. On this segment, the values of the signal are replaced with the linear imputation given by the following reassignment:\\[-11pt]
\begin{align}
\nonumber
\tilde \bx_{k,n}^{(i)}[m] \longleftarrow & \tilde \bx_{k,n}^{(i)}[m_1-1] \\
& \hspace{-9pt}+\! \dfrac{m\!-\!m_1\!\!+\!1}{m_2\!-\!m_1\!\!+\!2}\!\left( \tilde\bx_{k,n}^{(i)}[m_2\!\!+\!1] \!-\!\tilde\bx_{k,n}^{(i)}[m_1\!\!-\!1]\!\right)\!.
\end{align}

\subsubsection{Matching Pursuit}
\label{sse:MP}

Visual inspection and statistical exploration of the AU time courses indicated that most AU time courses are sparse.  AUs are typically active for brief periods with characteristic activity shapes.  High-frequency and low-amplitude changes typically represent model noise or incomplete facial movements.  We used the matching pursuit technique~\cite{Mallat1993} to remove uninformative weak variations in AU signals while preserving the most characteristic peaks. Matching pursuit optimally decomposes a given signal into a dictionary of basis functions using a minimal number of elements belonging to this dictionary, called atoms. The dictionary used to decompose the signals of the $n$th experiment is comprised of the following functions:
\begin{itemize}
\item the Gaussian window,
\begin{equation}
\bg_{\mu,\sigma}[m] = \exp\left(\dfrac{-1}{2\sigma^2}\left(m-\mu)\right)^2\right)\,,
\label{eq:gaussian}
\vspace{-2pt}
\end{equation}
where $\mu\in\{1,\ldots,M_n\}$, $\sigma\in\{\sigma_0,\ldots,\sigma_S\}$;
\item the Mexican hat wavelet,
\begin{equation}
\bw_{\mu,\sigma}[m] = \left(1-\dfrac{1}{\sigma^2}\left(m-\mu\right)^2\right)\bg_{\mu,\sigma}[m]\,,
\label{eq:mexican.hat}
\vspace{-1pt}
\end{equation}
where $\mu\in\{1,\ldots,M_n\}$, $\sigma\in\{\sigma_0,\ldots,\sigma_S\}$.
\end{itemize}
The dictionary thus contains $2M_n S$ atoms. The Gaussian-shaped atoms isolate peaks and bumps from the signal while the Mexican hat-shaped atoms capture areas of rapid change around the bumps.

We implemented the standard matching pursuit algorithm. Let $\hat{\bx}_{k,n}^{(i)}$ denote the output of the matching pursuit algorithm applied to the smoothed signal $\tilde \bx_{k,n}^{(i)}$. Then, $\hat{\bx}_{k,n}^{(i)}$ is the projection of $\tilde \bx_{k,n}^{(i)}$ onto a finite number $Q\ll 2M_n S$ of atoms that minimizes the squared distance $\|\hat{\bx}_{k,n}^{(i)}-\tilde{\bx}_{k,n}^{(i)}\|^2$.

\subsection{Step 2: Compute Interactional Synchrony}
\label{sse:DTW}
To assess interactional synchrony, or overall temporal coordination between AU time series pairs, we propose a new detection procedure based on DTW~\cite{berndt1994using} and its extensions. DTW estimates the function of local time shifts that minimizes the overall misfit between time series.  It does not assume any kind of stationarity in signals.  The DTW warping function describes how to shrink and stretch individual parts of each time series so that the resulting signals are maximally aligned.  By construction, ordinary DTW seeks an alignment $(u_{k,n}[t],v_{k,n}[t])_{t\in\{1,\ldots,T\}}$ of both
signals $\hat{\bx}_{k,n}^{(1)}$ and $\hat{\bx}_{k,n}^{(2)}$ that minimizes the following function:
\begin{equation}
D_{k,n} = \sum_{t=1}^T \left|\hat{\bx}_{k,n}^{(1)}[u_{k,n}[t]] - \hat{\bx}_{k,n}^{(2)}[v_{k,n}[t]]\right|.
\label{eq:dtw.distance}
\end{equation}
The following constraints on the warping path $(u_{k,n}[t], v_{k,n}[t])_{t\in\{1,\ldots,T\}}$ are applied to prevent the alignment from rewinding signals in time and to prevent signal samples from being omitted:
\begin{align}
u_{k,n}[t] \leq u_{k,n}[t+1]\,,&\quad v_{k,n}[t] \leq v_{k,n}[t+1]\,,\\
u_{k,n}[1] = v_{k,n}[1] = 1\,,&\quad u_{k,n}[T] = v_{k,n}[T] = M_n\,.
\end{align}

Since interactional synchrony is more related to the coordinated timing of movements than the coordination of movements' magnitudes, we implemented a version of DTW called \emph{\textit{derivative} DTW} (DDTW) \cite{keogh2001derivative}. DDTW estimates the function of local time shifts that minimizes the overall misfit between the \textit{derivatives} of a pair of time series instead of working on the raw time series. The functional result is that the alignment is influenced more by the time series' shapes than their magnitude.  DDTW is also more resilient than DTW to ``singularities'', or instances where a single point from one time series is mapped onto a large subsection of the other time series in an unintuitive manner.   
 
The DTW distance, or $D_{k,n}$ in equation~\eqref{eq:dtw.distance}, is typically used to assess similarity between two signals. $D_{k,n}$ represents the sum of the distances between corresponding points of the optimally warped time series \cite{hoch2021dancing, berndt1994using}.  Of note, although $D_{k,n}$ is referred to as a distance, it does not meet the mathematical definition of a distance because it does not guarantee the triangle inequality to hold.  When the DTW distance is used in the present study, it is normalized by the session's duration; that is, by the ratio $D_{k,n}/M_n$. Here, we introduce an assessment of interactional synchrony that leverages the shape of the DTW path instead of the DTW distance.  This new assessment is motivated by the aforementioned idea that behaviorally-relevant interactional synchrony is more about the coordinated timing of movements than it is about precise mimicry.  The DTW distance primarily provides information about the difference in shapes of two individuals' AU activity bouts.  The DTW path, on the other hand, primarily provides information about how much shifting in time is required to optimally align similar AU activity bouts.  Thus, the DTW path should be more relevant to ``the temporal linkage of nonverbal behavior'' than the DTW distance.  We focused specifically on the warping path's median deviation from the diagonal (WP-meddev). This quantity, denoted $\bz_n\in \RR^{K_\AU}$, reads:
\begin{equation}
\vspace{-4pt}
\bz_n[k] = \dfrac1{\sqrt{2}}\times\mathrm{median}\left(\left| v_{k,n}[t]-u_{k,n}[t] \right|\right)_{t\in\{1,\ldots,T\}} .
\label{eq:median.deviation}
\end{equation}
Intuitively, the less two time series are temporally aligned, the more warping will be required to optimally align them, and the more frequently the warping functions will have dramatic deviations from the diagnonal.  As a result, the warping function's median distance across a session will be longer.  We hypothesized that WP-meddev would be a better representation of interactional synchrony than the DTW distance, and therefore would also be more useful for predicting trust.

\subsection{Step 3: Prediction}
\label{sse:elastic.net}

A critical goal of this research is to develop a procedure that can select which of many highly-correlated social synchrony inputs are behaviorally-relevant in an interpretable way.  In service of this goal, we chose elastic net penalized regression to relate DTW features to H's choices in the Trust Game \cite{zou2005regularization}.  Penalized regression methods are robust in settings where a large number of features are examined relative to the number of data points.  Lasso and elastic net regression are two specific penalized strategies that also impose sparsity on the feature set, and unlike most black-box models, the features that are retained in their models can be interpreted straightforwardly as being informative for predicting the outcome measure.  When multiple features are both correlated with each other (as AUs are known to be) and correlated with the outcome variable, though, lasso regression will randomly retain only one of the correlated features. Elastic net, on the other hand, combines the lasso and ridge penalty functions so that it retains the set of features within correlated groups that maximize model performance, while still imposing enough sparsity to prevent overfitting. Its characteristics are therefore ideal for the present setting.  Let $\cD$ denote the deviance of the binomial logistic regression. Recall the regression problem:
\begin{align}
\nonumber
\left(\hat\bbeta,\hat\beta_0 \right) = \arg \min_{\substack{\bbeta\in\RR^{K_\AU} \\ \beta_0\in\RR}} & \sum_{n=1}^{N_\xp} \cD\left(\by[n],\bbeta\bz_n+\beta_0\right) \\[-8pt]
& \hspace{-3pt} + \lambda \left(\dfrac{1-\alpha}{2}\|\bbeta\|_2^2 + \alpha\|\bbeta\|_1 \right),
\label{eq:elastic.net}
\vspace{-4pt}
\end{align}
where $\lambda > 0$, and $\alpha\in[0,1]$ are hyperparameters. In experiments, $\lambda$ and $\alpha$ are chosen through a grid search that maximizes the accuracy of the resulting model. Five-fold cross validation is applied to validate the results.

\section{Results}
\label{se:results}

\subsection{Trust Game Outcomes}
\label{sse:binarization}
\begin{figure}[hbt!]
\centering
\includegraphics[width=.45\textwidth]{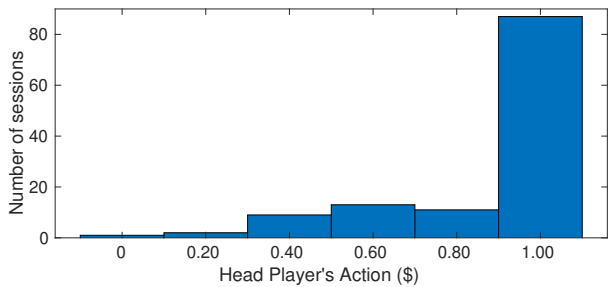}
\caption{Histogram of H player choices in the Trust Game.}
\label{fig:trustgame}
\end{figure}

To identify the interactional synchrony related to trust (as opposed to trustworthiness), we focused solely on H's actions.  The distribution of Hs' actions was highly unbalanced, since most H players chose to give the full \$1 (Fig.~\ref{fig:trustgame}).  To alleviate the statistical challenges of predicting such unbalanced classes, all subsequent analyses treat Trust Game behavior as a binary variable where trust class $0$ is associated with H choices ranging from \$0 through \$0.80 and trust class $1$ is associated with H choices of \$1.

\subsection{Smoothing and Matching Pursuit Preprocessing}

\begin{figure}[hbt!]
\centering
\includegraphics[width=.46\textwidth]{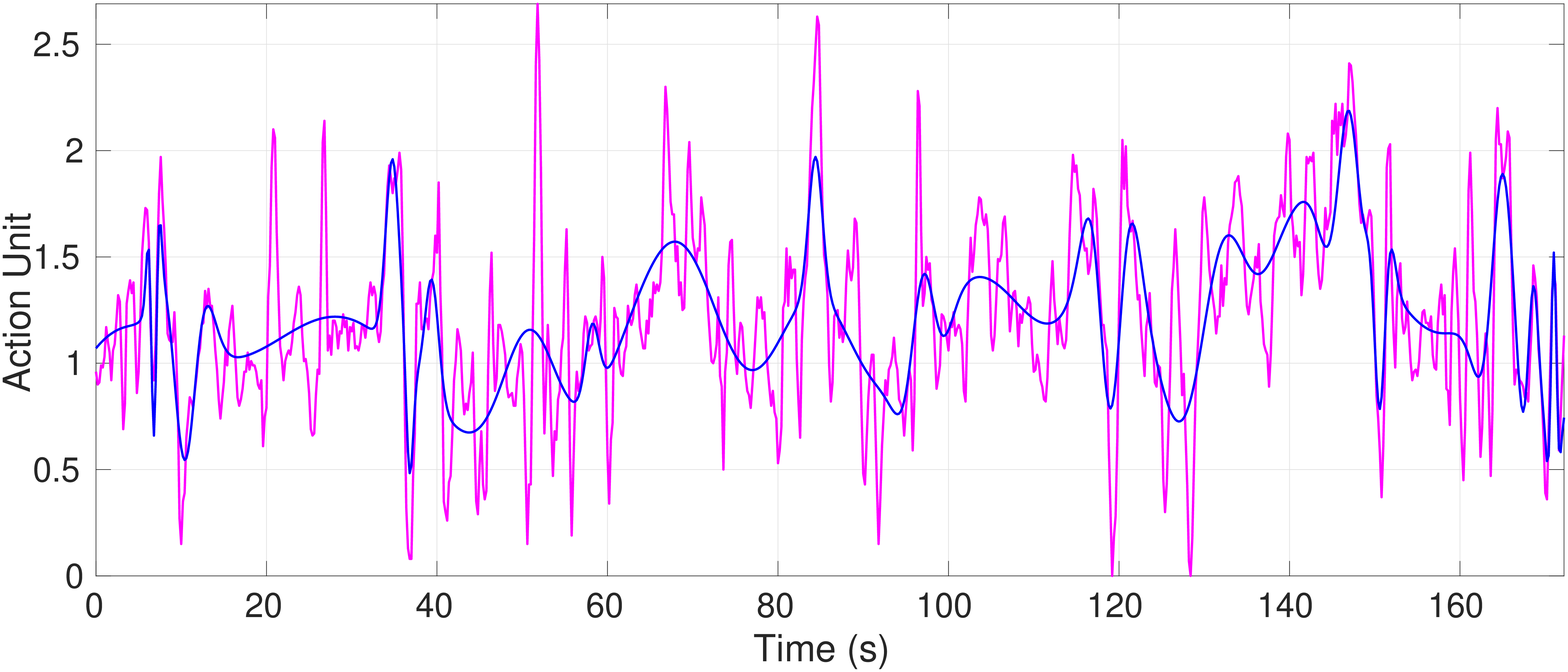}
\caption{Brow Lower AU signal of a participant before (in magenta) and after smoothing and matching pursuit preprocessing (in blue).}
\label{fig:AUsmoothingMP}
\end{figure}

Fig.~\ref{fig:AUsmoothingMP} depicts an example of a ``Brow Lower'' AU signal reconstructed after smoothing and matching pursuit preprocessing (depicted in blue). Matching Pursuit retains the most significant variations in the time series while removing small, random fluctuations. As little as 3\% and no more than 11\% of information was lost from each AU signal after matching pursuit (Supplementary Table~S-I).

\subsection{Dynamic Time Warping}

We hypothesized that DDTW would be better suited for assessing interactional synchrony than ordinary DTW (section~\ref{sse:DTW}), but compared the performance of both kinds of DTW.  We set $\Theta$, or the maximal time lag permitted when aligning signals (i.e., $|u_t-v_t|\leq\Theta/\fs$), to 5~s for our primary analyses described in subsequent sections, since psychology studies suggest interactional synchrony events occur with time lags of up to 5 seconds~\cite{delaherche2012interpersonal}. The results of alternative $\Theta$ values are provided in Section~II-A of the Supplementary Information, but consistent with the psychology literature, models with $\Theta$ values of 5~s performed the best.

\begin{figure}[hbt!]
\centering
\includegraphics[width=.45\textwidth]{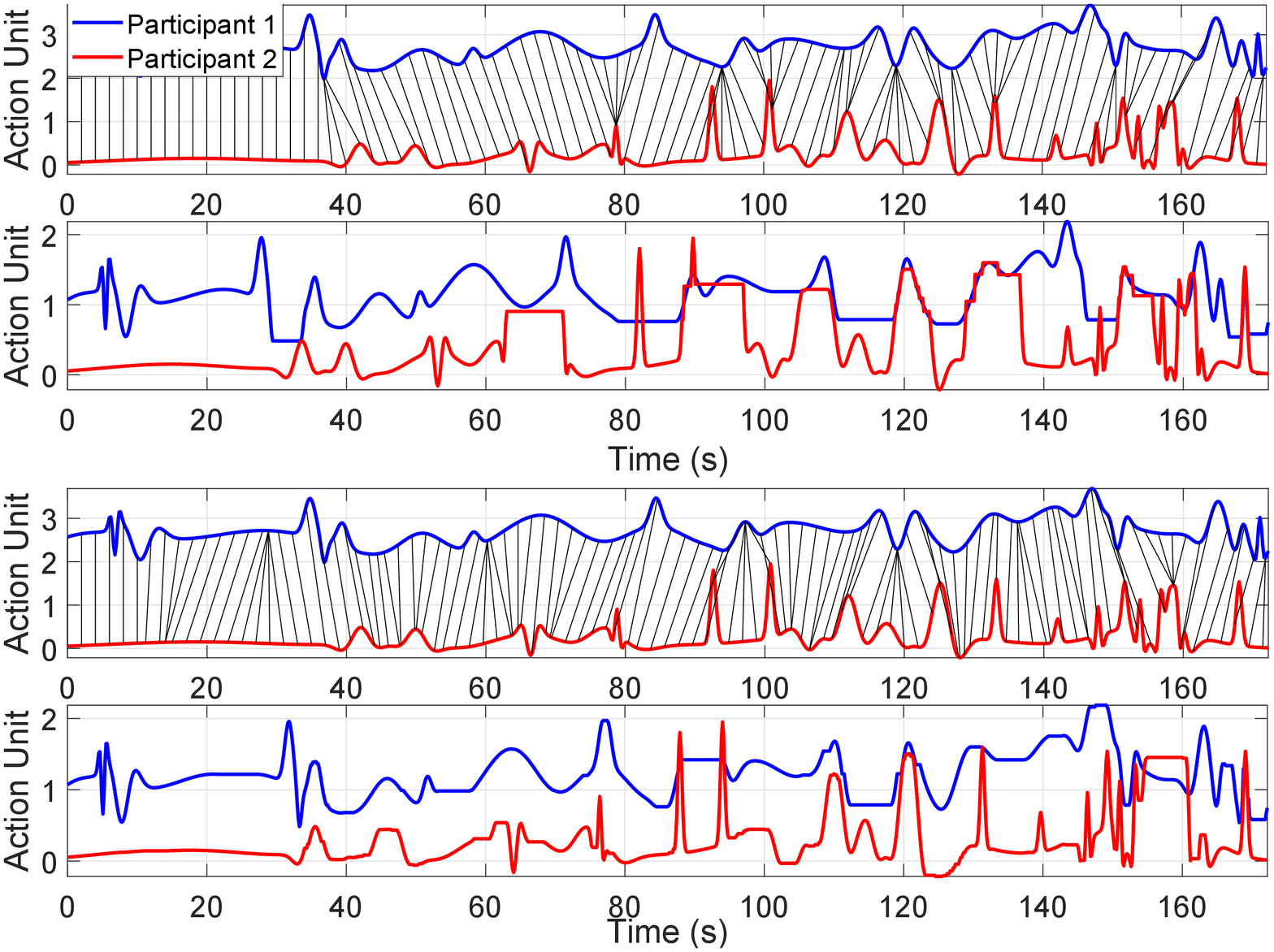}
\caption{Brow Lower AU time series of two participants. The black lines indicate which time points are aligned via ordinary DTW (first panel) vs. DDTW (third panel). The signals warped by the shifts prescribed by the optimal DTW (second panel) and DDTW (fourth panel) warping paths are shown below the alignments.  Blue and red indicate which participant the signal is from.}  
\label{fig:alignedAUs}
\end{figure}

\begin{figure}[hbt!]
\centering
\includegraphics[width=.48\textwidth]{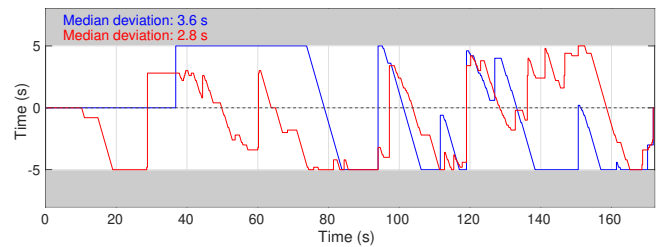}
\caption{Deviation from the warping path diagonal obtained via DTW (blue) and DDTW (red) applied to the AU time series displayed in Fig.~\ref{fig:alignedAUs}.}
\label{fig:warping.path}
\end{figure}

An example pair of AUs aligned by DTW vs. DDTW is shown in Fig.~\ref{fig:alignedAUs}. The benefits of DDTW are apparent. DDTW avoids some of DTW's unrealistic alignments where a single point within a peak of one signal is inappropriately matched to a stretched segment of the other signal that is artificially made to be uniformly flat (e.g. the segments between 20--40s and 80--100s). Fig.~\ref{fig:warping.path} shows the deviation from the diagonal of the warping paths obtained via DDTW vs DTW, and their associated  WP-meddev values. The constraints imposed by $\Theta$ are depicted in gray.  Departures from the diagonal indicate alignments of samples initially distant in time (e.g., the segment between 80--100s).  

\subsection{Prediction Procedure}

We evaluated the ability of multivariate interactional synchrony, as measured by univariate AU's median deviation from the diagonal of the DDTW's warping path (WP-meddev), to predict the outcome of the Trust Game. Even with the binary transformation of H's behavior detailed in section~\ref{sse:binarization}, trust behavior represented by the variable $\by$ remained imbalanced. The number of sessions belonging to trust class $0$ was $N_0=36$, while the number of sessions belonging to trust class $1$ was $N_1=87$. To perform prediction, we randomly subsampled the overrepresented class so that only $36$ sessions belonging to the trust class $1$ were retained. The total number of sessions included in the subsequent prediction analyses were therefore $72$, equally balanced between trust behavior classes $0$ and $1$. 

The prediction problem was solved via the elastic net procedure introduced in section~\ref{sse:elastic.net} as follows.  The data set was partitioned into five subsamples.  The parameters $(\hat\bbeta,\hat\beta_0)$ were learned from a training set (about 58 sessions) comprised of four subsamples, and then tested by predicting the Trust Game outcomes in the testing set (about 14 sessions) comprised of the fifth subsample.  This was repeated with different subsamplings of trust class \textbf{1} until each session had been considered at least 50 times.

The above procedure was run on a grid of possible values for the elastic net model's hyperparameters $\lambda$ and $\alpha$.  Values of $\lambda=0.0518$ and $\alpha=0.802$ maximized the accuracy of the models applied to the non-imputed data. Since the prediction accuracy did not change much with a wide range of hyperparameter values (see Section~II-B of Supplementary Information), the identified $\lambda$ and $\alpha$ values were retained for all subsequent analyses. Table~\ref{tab:accuracy} shows how often the WP-meddev models (indicated by ``WP'' in the table) correctly predicted Trust Game outcomes when applied to non-imputed AU signals or AU signals whose low-confidence frames were linearly imputed (see section~\ref{sse:imputation}).  The accuracy rates of WP-meddev models were 63.4--67.7\%, compared to the 50\% that would be expected by chance.  These results indicate that interactional synchrony between AUs is indeed informative for predicting trust.

\begin{table}[hbt!]
\centering
\caption{Prediction Accuracy, obtained via successive 5-fold cross validations that preserve the class distribution.}

\begin{tabular}{|p{40pt}|p{30pt}|p{38pt}||p{22pt}|p{22pt}|p{22pt}|}
\hline
\multirow{2}{*}{\bfseries{Model Name}} &
\multirow{2}{*}{\bfseries{Measure}} & \multirow{2}{45pt}{\bfseries{Features Extracted}} & \multicolumn{3}{c|}{\bfseries{Elastic Net}} \\ \cline{4-6}
& &  & \scriptsize\textbf{Class 0} & \scriptsize\bfseries{Class 1}  & \scriptsize\bfseries{Overall} \\ \hhline{|=|=|=#=|=|=|} 
WP-DDTW & WP-meddev & Nonimputed AUs & $\mathbf{63.4\%}$ & $\mathbf{65.0\%}$ & $\mathbf{64.2\%}$\\ \hline
WP-DDTW (imputed) & WP-meddev & Imputed AUs & $\mathbf{67.5\%}$ & $\mathbf{67.7\%}$ & $\mathbf{67.6\%}$ \\ \hline
WP-DTW & WP-meddev & Nonimputed AUs & $52.5\%$ & $53.2\%$ & $52.9\%$ \\ \hline
WP-DTW (imputed) & WP-meddev & Imputed AUs & $48.3\%$ & $46.5\%$ & $47.4\%$ \\ \hline
\hline
\multicolumn{6}{|c|}{\bfseries{Shuffled Data Sets}} \\
\hline
Shuffled Pairs & WP-meddev & Nonimputed AUs  & $48.6\%$ & $50.1\%$ & $49.3\%$ \\ \hline
Shuffled Time Series & WP-meddev & Nonimputed AUs  & $51.1\%$ & $49.2\%$ & $50.1\%$ \\ \hline
\hline
\multicolumn{6}{|c|}{\bfseries{WCC Models}} \\
\hline
WCC-MEA & \scriptsize WCC (duration) & MEA & $54.2\%$ & $55.0\%$ & $54.6\%$ \\ \hline
WCC-AUs & \scriptsize WCC (duration) & Nonimputed AUs & $55.3\%$ & $57.4\%$ & $56.4\%$ \\ \hline
\hline
\multicolumn{6}{|c|}{\bfseries{Univariate WP-MEA Model}} \\
\hline
WP-MEA & WP-meddev & MEA & $25.8\%$ & $68.7\%$ & $47.3\%$ \\ \hline
\hline
\multicolumn{6}{|c|}{\bfseries{DTW Distance Models}} \\
\hline
DDTW distance & DDTW distance & Nonimputed AUs & $45.1\%$ & $47.9\%$ & $46.5\%$ \\ \hline
DTW distance & DTW distance & Nonimputed AUs & $51.2\%$ & $57.0\%$ & $50.3\%$ \\ \hline
\hline
\multicolumn{6}{|c|}{\bfseries{Optimal Transport Model}} \\
\hline
Optimal transport & EMD & Nonimputed AUs & $55.3\%$ & $57.2\%$ & $56.3\%$ \\ \hline
\hline
\multicolumn{6}{|c|}{\bfseries{Multivariate Models with independent AU Features (no synchrony)}} \\
\hline
\scriptsize AU-Durations (H) & AUs durations & Nonimputed AUs & $44.8\%$ & $49.8\%$ & $47.3\%$ \\ \hline
\scriptsize AU-Durations (T) & AUs durations & Nonimputed AUs & $50.9\%$ & $55.7\%$ & $53.3\%$ \\ \hline
AU Intensities (H) & AUs intensities & Nonimputed AUs & $53.4\%$ & $56.1\%$ & $54.7\%$ \\ \hline
AU Intensities (T) & AUs intensities & Nonimputed AUs & $52.3\%$ & $51.4\%$ & $51.8\%$ \\ \hline
\end{tabular}

\label{tab:accuracy}
\end{table}

To ensure that the WP-meddev measure represents biologically meaningful signal, we computed WP-meddev for two control data sets:
\begin{enumerate}
\item \emph{Shuffled pairs.} Videos were randomly shuffled so that each H player video was paired with another video from a session of the same trust class, but where the paired partners did not actually interact with each other over Skype.  Any synchrony between these videos would be due to chance rather than due to natural synchrony between interacting partners.
\item \emph{Shuffled time series.} All video pairs from the same session were divided into 10-second intervals, and these 10-second intervals were randomly shuffled. The same shuffling was applied to all AU time series. If WP-meddev tracks true temporal coordination between AU time series, the shuffling procedure should disrupt the accurate assessment of interactional synchrony.
\end{enumerate}

The accuracy of the WP-meddev prediction models using these two control data sets was worse than chance (Table~\ref{tab:accuracy}).  This confirms that the WP-meddev interactional synchrony measure tracks real dynamics between interacting human partners rather than just coincidental temporal coordination that can be expected from any random pair of time series.

Next, we assessed the predictive utility of WP-meddev compared to other features that might be extracted from social interaction videos.  We began by comparing the predictive utility of WP-meddev to the MEA WCC-duration method described in section~\ref{se:Introduction} (WCC-UV, where ``UV'' represents a univariate head region predictor) using the parameters recommended by Altmann~\cite{Altmann2011}. Table~\ref{tab:accuracy} shows that WCC-UV leads to an accuracy of around 55\%, which is better than chance, but worse than our WP-meddev method.  To determine whether the reduction in prediction accuracy is due to the measure used to assess interactional synchrony or, instead, due to treating the entire head region as a single feature, we ran two additional analyses. The first used the multivarate elastic net procedure described above with the AU time courses, but used WCC instead of WP-meddev to assess interactional synchrony (WCC-AUs).  The second examined the univariate relationship between the MEA time series and trust, but used WP-meddev instead of WCC to assess interactional synchrony (WP-MEA).  Table~\ref{tab:accuracy} shows that the multivariate WP model outperformed the WCC-AUs model, confirming that WP-meddev is a more informative interactional synchrony measure than WCC in this context.  The multivariate WP-meddev model also outperformed the WP-MEA model, indicating that examining more fine grained interactional synchrony between AUs is more informative for predicting trust than examining interactional synchrony between movement in the head region as a whole.  

Next we tested the predictive utility of the DTW distance between each AU pair, since similarity is often assessed using DTW distance.  As discussed in \ref{sse:DTW}, the DTW distance is the sum of the normalized euclidean distances between corresponding points of optimally warped time series, and is a fundamentally different measure than the WP-meddev measure we have introduced.  In illustration, the Pearson correlation coefficient between the DTW/DDTW distances and WP-meddev measures of all AU pairs in the current data set is $0.24$ ($\text{p}<0.001$) and $-0.19$ ($\text{p}<0.001$), respectively.  This indicates the relationship is not only small, but in the case of DDTW, also in the negative direction.  We ran one elastic net model using the DTW distance of AU pairs as features and another using the DDTW distance of AU pairs as features (both were normalized by the duration of the session).   The accuracy of both models was poor, and in the case of the DDTW distance, was worse than chance.  This confirmed our prediction that the WP-meddev interactional synchrony measure would be more behaviorally-relevant than traditional DTW distance measures.

We also tested the predictive utility of another popular similarity measure, the optimal transport distance. Optimal transport approaches calculate the cost of moving one distribution of data to another, taking spatial proximity into account \cite{peyre2019computational}.  They cannot assess the temporal coordination between two time series because they treat each time point as a member of a collection where chronological order is ignored, so conceptually, they are not well-suited to index interactional synchrony.  Yet, we can take advantage of the fact that they effectively assess the similarity of the magnitudes of two time-series, even when similar magnitudes are shifted in time. We tested the earth mover's distance (EMD), the most common transport distance. The elastic net models using the EMDs between AU pairs as features performed similarly to MEA-WCC models.  Both types of models predicted trust much less successfully than WP-meddev models, providing converging evidence that the temporal coordination between AUs plays a unique role in predicting trust, beyond information provided by coordination of AU magnitudes. 

To test whether WP-meddev interactional synchrony features were more informative for predicting trust than features of the multivariate AU time series from each player considered independently, we examined the performance of models that used the duration of AU features demonstrated by the H and T players as features (AU durations in Table~\ref{tab:accuracy}), and models that used the average  intensity of the H and T players' AUs across a session as features (AU intensities in Table~\ref{tab:accuracy}).  These features are similar to those used by previous studies trying to predict trust using automatic visual feature detection~\cite{lucas2016trust}. The duration of an AU was defined by the proportion of time the AU was detected as visible (AU intensity $> 1$) in the H or T player, considered separately.  Table~\ref{tab:accuracy} shows that the AU-Durations models and AU-Intensities models underperformed relative to most of the interactional synchrony models.  The AU-Intensities model from the H player had the best performance of the four, but was still much less accurate than the WP-DDTW models.  This confirms that extracting information about how the facial features of a pair of people interact with each other over time is generally more helpful for predicting trust than extracting information about the people's facial features considered independently from one another.

Finally, we compared the performance of all the elastic net models to the accuracy of random forest models using the same features and behavioral labels~\cite{Breiman2001} (see Supplementary Information for model details).  In the present study, the elastic net procedure always outperformed the random forest models (see Table~S-III in Supplementary Information).  This suggests the elastic net strategy is better suited for identifying the specific types of interactional synchrony that predict trust or other types of behaviors of interest. That said, the fact that the performance of both algorithms were similar suggests that the relatively modest 60--65\% accuracy rate of the models likely reflects an imperfect relationship between interactional synchrony predictors and trust more than an unsuitable modeling strategy or inappropriate statistical assumptions.

\begin{figure}[hbt!]
\centering
\includegraphics[width=.48\textwidth]{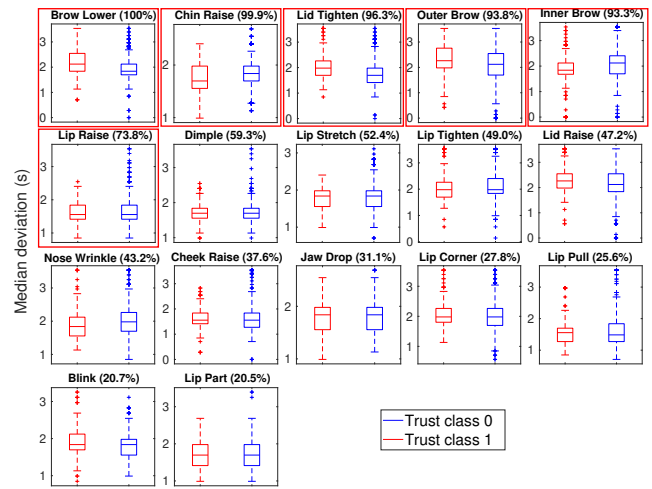}
\caption{Box plots of the DDTW warping paths' median deviation from the diagonal for each trust class. The percent of experiments where an AU was retained in the elastic net model are provided in parentheses. The red boxes indicate when this percentage is greater than 70\%.}
\label{fig:boxplotAUs}
\end{figure}

Taking advantage of the feature selection built into elastic net models, Fig. ~\ref{fig:boxplotAUs} displays the percent of experiments where an AU was retained in the elastic net model (meaning the estimated parameter vector $\hat\bbeta$ for the AU was nonzero).  The most frequently retained AUs are the AUs that are most informative for predicting trust.  It is notable that four of the six AUs that were selected by more than 70\% of the models are eye-related---Brow Lower, Lid Tighten, Outer Brow and Inner Brow (Blink and Lid Raise are the only eye-related AUs that are not selected regularly).  Fig.~\ref{fig:boxplotAUs} displays the box plots of each AU's WP-meddev interactional synchrony according to the outcome of the Trust Game. The AUs with greater interactional synchrony differences between the two trust classes are retained in a greater percentage of experiments, confirming the linear models are sensible.

\section{Conclusion and Perspective}

We have demonstrated that automatic analysis of interactional synchrony during unconstrained social interactions can be used to predict trust in a subsequent Trust Game.  We also described a procedure that identifies which interactional synchrony features from a multivariate set are behaviorally relevant.  Three overarching conclusions can be drawn from this work.  First, analyzing the temporal interactions between people provides unique insight into social behavior that cannot be gleaned by analyzing the partners actions in isolation.  Second, the median deviation of DDTW warping paths may be a more effective way of studying these interactions than any other interaction measure previously described.  Third, multivariate approaches to studying interactional synchrony may be more fruitful than univariate approaches.

Of course, this study has limitations.  Most notably, although our data set is one of the largest of its kind, it still only contained 123 usable sessions and its behavioral data was highly unbalanced.  Thus, no strong conclusions should be made about the AUs found to predict trust in the current study until the presented analyses can be tested on additional data.  There are also ways the presented analytical strategy could be improved.  The WP-meddev elastic net models we designed performed better than any other model tested, but their accuracy might be enhanced in the future by including additional movement or facial features, finding better representations of those movements than are provided by facial action units, or relaxing the interpretability requirements so that a greater variety of prediction algorithms can be employed (like DNN).  Further, parts of the proposed analytical method can be further optimized and automated.  In particular, we limited the time lags that could be imposed on the AU time series to 5 seconds (represented by $\Theta$), because that was the time lag analyzed in most previous interactional synchrony studies.  Our exploratory sensitivity analysis confirmed that a $\Theta$ value of 5 led to most accurate model of those tested, but establishing a method for automating the selection of $\Theta$ in a data-driven fashion would greatly benefit other applications of the method.  Another issue that requires attention is that the quality of feature extraction from videos can impact synchrony detection, especially when signal cleaning and preprocessing is not optimized.  Our prediction results improved when frames with low confidence values were imputed, but more research is needed to determine which time points in a feature time series should be imputed and what imputation method will be most successful.

Notwithstanding these limitations, the methods described here provide new procedures and a novel interactional synchrony measure for identifying what kinds of multivariate facial and gesture interactional synchrony are important for social behaviors.  The information gleaned from applications of this work may be useful to many fields, and can be leveraged to understand psychiatric disease, develop more effective virtual agents, and create interventions that teach people how to have more successful social interactions.

{\small
\bibliographystyle{ieeetr}
\bibliography{RefsSocialSynchrony}

\begin{thebibliography}{10}

\bibitem{vicaria2016meta}
I.~M. Vicaria and L.~Dickens, ``Meta-analyses of the intra-and interpersonal
  outcomes of interpersonal coordination,'' {\em J. Nonverbal Behav.}, vol.~40,
  no.~4, pp.~335--361, 2016.

\bibitem{schirmer2021being}
A.~Schirmer, M.~Fairhurst, and S.~Hoehl, ``{Being ‘in sync’—is
  interactional synchrony the key to understanding the social brain?},'' {\em
  Soc. Cogn. Affect. Neurosci.}, vol.~16, pp.~1--4, 10 2020.

\bibitem{rennung2016prosocial}
M.~Rennung and A.~S. G{\"o}ritz, ``Prosocial consequences of interpersonal
  synchrony,'' {\em Zeitschrift f{\"u}r Psychologie}, 2016.

\bibitem{delaherche2012interpersonal}
E.~Delaherche, M.~Chetouani, A.~Mahdhaoui, C.~Saint-Georges, S.~Viaux, and
  D.~Cohen, ``Interpersonal synchrony: A survey of evaluation methods across
  disciplines,'' {\em IEEE Trans. Affective Comput.}, vol.~3, no.~3,
  pp.~349--365, 2012.

\bibitem{vinciarelli2009social}
A.~Vinciarelli, H.~Salamin, and M.~Pantic, ``Social signal processing:
  Understanding social interactions through nonverbal behavior analysis,'' in
  {\em 2009 IEEE Comput. Soc. Conf. Comput. Vis. Pattern Recognit. Workshops},
  pp.~42--49, 2009.

\bibitem{kruzic2020facial}
C.~O. Kruzic, D.~Kruzic, F.~Herrera, and J.~Bailenson, ``Facial expressions
  contribute more than body movements to conversational outcomes in
  avatar-mediated virtual environments,'' {\em Scientific Reports}, vol.~10,
  no.~1, pp.~1--23, 2020.

\bibitem{goldstein2020clinician}
P.~Goldstein, E.~A.~R. Losin, S.~R. Anderson, V.~R. Schelkun, and T.~D. Wager,
  ``Clinician-patient movement synchrony mediates social group effects on
  interpersonal trust and perceived pain,'' {\em The Journal of Pain}, vol.~21,
  no.~11-12, pp.~1160--1174, 2020.

\bibitem{boker2002windowed}
S.~M. Boker, J.~L. Rotondo, M.~Xu, and K.~King, ``Windowed cross-correlation
  and peak picking for the analysis of variability in the association between
  behavioral time series.,'' {\em Psychological methods}, vol.~7, no.~3,
  p.~338, 2002.

\bibitem{berg1995trust}
J.~Berg, J.~Dickhaut, and K.~McCabe, ``Trust, reciprocity, and social
  history,'' {\em Games Econ. Behav.}, vol.~10, no.~1, pp.~122--142, 1995.

\bibitem{todorov2009evaluating}
A.~Todorov, M.~Pakrashi, and N.~N. Oosterhof, ``Evaluating faces on
  trustworthiness after minimal time exposure,'' {\em Social Cognition},
  vol.~27, no.~6, pp.~813--833, 2009.

\bibitem{bonnefon2017can}
J.-F. Bonnefon, A.~Hopfensitz, and W.~De~Neys, ``Can we detect cooperators by
  looking at their face?,'' {\em Current Directions in Psychological Science},
  vol.~26, no.~3, pp.~276--281, 2017.

\bibitem{barrett2019emotional}
L.~F. Barrett, R.~Adolphs, S.~Marsella, A.~M. Martinez, and S.~D. Pollak,
  ``Emotional expressions reconsidered: Challenges to inferring emotion from
  human facial movements,'' {\em Psychological science in the public interest},
  vol.~20, no.~1, pp.~1--68, 2019.

\bibitem{ekman1997face}
P.~Ekman and E.~L. Rosenberg, {\em What the face reveals: Basic and applied
  studies of spontaneous expression using the Facial Action Coding System
  (FACS)}.
\newblock Oxford University Press, USA, 1997.

\bibitem{baltrusaitis2018openface}
T.~Baltrusaitis, A.~Zadeh, Y.~C. Lim, and L.-P. Morency, ``Open{F}ace 2.0:
  Facial behavior analysis toolkit,'' in {\em 2018 13th IEEE Int. Conf. Autom.
  Face Gesture Recognit. (FG 2018)}, pp.~59--66, 2018.

\bibitem{Mallat1993}
S.~Mallat and Z.~Zhang, ``Matching pursuits with time-frequency dictionaries,''
  {\em IEEE Trans. Signal Process}, vol.~41, no.~12, pp.~3397--3415, 1993.

\bibitem{berndt1994using}
D.~J. Berndt and J.~Clifford, ``Using dynamic time warping to find patterns in
  time series,'' in {\em Proc. 3rd Int. Conf. Knowl. Discov. Data Min.},
  AAAIWS'94, pp.~359--370, AAAI Press, 1994.

\bibitem{keogh2001derivative}
E.~J. Keogh and M.~J. Pazzani, ``Derivative dynamic time warping,'' in {\em
  Proc. 2001 SIAM Int. Conf. Data Min.}, pp.~1--11, SIAM, 2001.

\bibitem{hoch2021dancing}
J.~E. Hoch, O.~Ossmy, W.~G. Cole, S.~Hasan, and K.~E. Adolph, ``“{D}ancing”
  together: Infant--mother locomotor synchrony,'' {\em Child development},
  2021.

\bibitem{zou2005regularization}
H.~Zou and T.~Hastie, ``Regularization and variable selection via the elastic
  net,'' {\em J. R. Stat. Soc. Series B Stat. Methodol.}, vol.~67, no.~2,
  pp.~301--320, 2005.

\bibitem{Altmann2011}
U.~Altmann, ``Investigation of movement synchrony using windowed cross-lagged
  regression,'' in {\em Analysis of Verbal and Nonverbal Communication and
  Enactment. The Processing Issues}, pp.~335--345, 2011.

\bibitem{peyre2019computational}
G.~Peyr{\'e} and M.~Cuturi, ``Computational optimal transport: With
  applications to data science,'' {\em Foundations and Trends{\textregistered}
  in Machine Learning}, vol.~11, no.~5-6, pp.~355--607, 2019.

\bibitem{lucas2016trust}
G.~Lucas, G.~Stratou, S.~Lieblich, and J.~Gratch, ``Trust me: multimodal
  signals of trustworthiness,'' in {\em Proc. 18th ACM Int. Conf. Multimodal
  Interact.}, pp.~5--12, 2016.

\bibitem{Breiman2001}
L.~Breiman, ``Random forests,'' {\em Machine Learning}, vol.~45, pp.~5--321,
  2001.

\end{thebibliography}
}

\end{document}


\ifFGfinal
\thispagestyle{empty}
\pagestyle{empty}
\else
\author{Anonymous FG2021 submission\\ Paper ID \FGPaperID \\}
\pagestyle{plain}
\fi
\maketitle

\section{Effect of Matching Pursuit Preprocessing on the AUs}

We performed the smoothing and matching pursuit steps described in section~III-C.1 and section~III-C.3 of the main document on all 17 AU signals. Matching pursuit was performed on the smoothed signals.  We set the number of selected atoms to $Q=25$ per signal. The atom shapes are described in section~III-C.3, where $\sigma\in\{2^1,\ldots,2^5\}$.

Let $\hat{\bx}_{k,n}^{(i)}$ denote the preprocessed version of the original signal $\bx_{k,n}^{(i)}$. The measure of the relative amount of information lost in this step is determined by
\begin{align*}
\ell_k = \frac{\ell_k^{(1)}+ \ell_k^{(2)}}{2} \quad\mathrm{where}\quad
\ell_k^{(i)} = \dfrac1{N_\xp} \sum_{n=1}^{N_\xp} \dfrac{\|\hat{\bx}_{k,n}^{(i)} - \bx_{k,n}^{(i)} \|^2}{\|\bx_{k,n}^{(i)}\|^2} ,
\end{align*} 
for all $k\in\{1,\ldots,K_\AU\}$.
Table~\ref{tab:lost} shows the loss quantity $\ell_k$ for the 17 AUs that are available through OpenFace. Matching Pursuit was able to recover most of the information in the AU time courses, with no information loss exceeding 11.19\%.  The greatest information loss was from the blink signal time course, perhaps because it had more and faster overall variability than other AUs that were more sparse. 

\begin{table}[hbt!]
\centering
\caption{Percent information lost from each AU signal by matching pursuit pre-processing}
\begin{tabular}{lc}
\hline
Action Unit  & Loss (\%) \\
\hhline{==}
Blink  & 11.17 \\
Lip Stretch  & 7.79 \\
Lip Corner  & 7.49 \\
Nose Wrinkle  & 7.19 \\
Jaw Drop  & 7.14 \\
Brow Lower & 6.96 \\
Chin Raise  & 6.77 \\
Lid Tighten  & 5.87 \\
Lid Raise  & 5.84 \\
Lip Tighten  & 5.81 \\
Inner Brow  & 5.42 \\
Outer Brow  & 5.13 \\
Lip Raise  & 4.62 \\
Dimple  & 4.45 \\
Lip Part  & 4.27 \\
Cheek Raise  & 3.61 \\
Lip Pull  & 3.23 \\
\hline
\end{tabular}
\label{tab:lost}
\end{table}

\section{Influence of DDTW and Elastic Net Parameters on Model Prediction Accuracy}

\subsection{Influence of DDTW Time Lag Limit}
The parameter $\Theta$ represents the maximal time lag permitted when aligning pairs of signals via dynamic time warping. As described in the main text, we chose a $\Theta$ value of 5 seconds for the main experiments \emph{a priori} to ensure our analyses were consistent with the time lags permitted in the psychological interactional synchrony literature.  However, we took advantage of our automated procedure to test whether the lags most frequently used in the psychology literature are optimal.  The results indicated that the $\Theta$ choice does indeed heavily impact the performance of the prediction procedure. Table~\ref{tab:effect.theta} provides the overall prediction of the elastic net model applied to the nonimputed AUs for different values of DDTW $\Theta$. The value $\Theta=5$~s maximizes the prediction accuracy, thus allowing an effective measurement of the interactional synchrony.  

\begin{table}[hbt!]
\centering
\caption{Elastic net model accuracy for different values of DDTW $\Theta$.}
\begin{tabular}{cc}
\hline
$\Theta$ (s) & Overall Accuracy (\%) \\
\hhline{==}
2 & 55.6 \\
3 & 55.7 \\
5 & 64.2 \\
10 & 56.8 \\
15 & 59.2 \\
20 & 61.2 \\
\hline
\end{tabular}
\label{tab:effect.theta}
\end{table}

\subsection{Influence of the Elastic Net Tuning Parameters}
\begin{figure}
\centering
\includegraphics[width=.48\textwidth]{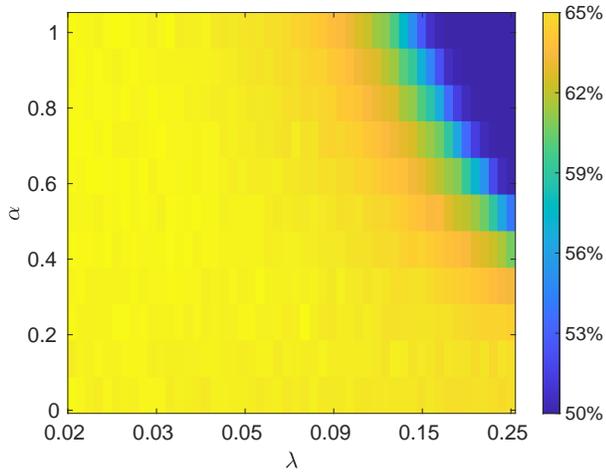}
\caption{Overall accuracy of the elastic net model across different $\alpha$ and $\lambda$ parameter values.}
\label{fig:influence.parameters}
\end{figure}

We evaluated the influence of changes in the $\lambda$ and $\alpha$ elastic net tuning parameters on prediction accuracy. To this end, the elastic net algorithm was run on the nonimputed AUs for a grid of possible tuning values, namely $\lambda\in[0.02,0.25]$ and $\alpha\in[0.01,1]$. Fig.~\ref{fig:influence.parameters} shows the overall accuracy according to the possible pairs of values. Except when $\lambda > 0.10$ and $\alpha > 0.5$ simultaneously, the overall accuracy remains around 65\%. Since changes in $\lambda$ and $\alpha$ values within these boundaries did not meaningfully impact performance, the  optimal $\lambda=0.0518$ and $\alpha=0.802$ values for this model were retained for all subsequent analyses.

\section{Comparison Between Elastic Net and Random Forest Models}

We compared the prediction accuracies of elastic net models with the accuracies of random forest models with the same inputs. The random forest prediction results from the bagging of 20 bootstrap-aggregated classification trees. Table~\ref{tab:accuracy} shows that elastic net outperforms random forest for most of the models that were tested.

\begin{table*}[b]
\centering
\caption{Prediction Accuracy, obtained via successive 5-fold cross validations that preserve the class distribution.}
\begin{tabular}{|c|c|c||c|c|c||c|c|c|}
\hline
\multirow{2}{*}{\bfseries{Model name}} &
\multirow{2}{*}{\bfseries{Measure}} & \multirow{2}{*}{\bfseries{Features extracted}} & \multicolumn{3}{c||}{\bfseries{Elastic Net}} & \multicolumn{3}{c|}{\bfseries{Random Forest}} \\ \cline{4-9}
& &  & \bfseries{Class 0} & \bfseries{Class 1}  & \bfseries{Overall} & \bfseries{Class 0} & \bfseries{Class 1} & \bfseries{Overall} \\ \hhline{|=|=|=#=|=|=#=|=|=|} 
WP-DDTW & WP-meddev & Non-imputed AUs & $\mathbf{63.4\%}$ & $\mathbf{65.0\%}$ & $\mathbf{64.2\%}$ & $67.8\%$ & $55.8\%$ & $61.8\%$  \\ \hline
WP-DDTW (imputed) & WP-meddev & Imputed AUs & $\mathbf{67.5\%}$ & $\mathbf{67.7\%}$ & $\mathbf{67.6\%}$ & $69.0\%$& $60.9\%$ & $65.0\%$ \\ \hline
WP-DTW & WP-meddev & Non-imputed AUs & $52.5\%$ & $53.2\%$ & $52.9\%$ & $59.5\%$  & $53.2\%$ & $56.4\%$ \\ \hline
WP-DTW (imputed) & WP-meddev & Imputed AUs & $48.3\%$ & $46.5\%$ & $47.4\%$ & $56.0\%$ & $48.8\%$ & $52.4\%$ \\ \hline
\hline
\multicolumn{9}{|c|}{\bfseries{Shuffled Data Sets}} \\
\hline
Shuffled Pairs & WP-meddev & Non-imputed AUs  & $48.6\%$ & $50.1\%$ & $49.3\%$ & $52.1\%$ & $46.6\%$ & $49.3\%$ \\ \hline
Shuffled Time Series & WP-meddev & Non-imputed AUs  & $51.1\%$ & $49.2\%$ & $50.1\%$ & $57.4\%$ & $50.0\%$ & $53.7\%$ \\ \hline
\hline
\multicolumn{9}{|c|}{\bfseries{WCC Models}} \\
\hline
WCC-MEA & WCC (duration) & MEA & $54.2\%$ & $55.0\%$ & $54.6\%$ & $53.5\%$ & $53.3\%$ & $53.4\%$\\ \hline
WCC-AUs & WCC (duration) & Non-imputed AUs & $55.3\%$ & $57.4\%$ & $56.4\%$ & $55.3\%$ & $50.4\%$ & $52.9\%$ \\ \hline
\hline
\multicolumn{9}{|c|}{\bfseries{Univariate WP-MEA Model}} \\
\hline
WP-MEA & WP-meddev & MEA & $25.8\%$ & $68.7\%$ & $47.3\%$ & $48.0\%$ & $41.7\%$ & $44.8\%$ \\ \hline
\hline
\multicolumn{9}{|c|}{\bfseries{DTW Distance Models}} \\
\hline
DDTW distance & DDTW distance & Non-imputed AUs & $45.1\%$ & $47.9\%$ & $46.5\%$ & $54.9\%$ & $45.7\%$ & $50.3\%$\\ \hline
DTW distance & DTW distance & Non-imputed AUs & $51.2\%$ & $57.0\%$ & $50.3\%$& $56.3\%$ & $49.2\%$ & $52.8\%$ \\ \hline
\hline
\multicolumn{9}{|c|}{\bfseries{Optimal Transport Model}} \\
\hline
Optimal transport & EMD & Non-imputed AUs & $55.3\%$ & $57.2\%$ & $56.3\%$ & $56.8\%$ & $54.3\%$ & $55.6\%$ \\ \hline
\hline
\multicolumn{9}{|c|}{\bfseries{Multivariate Models with independent AU Features (no interactional synchrony)}} \\
\hline
AU-Durations (H) & AUs durations & Non-imputed AUs & $44.8\%$ & $49.8\%$ & $47.3\%$ & $47.7\%$ & $42.3\%$ & $45.0\%$ \\ \hline
AU-Durations (T) & AUs durations & Non-imputed AUs & $50.9\%$ & $55.7\%$ & $53.3\%$ & $56.8\%$ & $49.3\%$ & $53.1\%$ \\ \hline
AU-Intensities (H) & AUs intensities & Non-imputed AUs & $53.4\%$ & $56.1\%$ & $54.7\%$ & $56.8\%$ & $49.1\%$ & $53.0\%$ \\ \hline
AU-Intensities (T) & AUs intensities & Non-imputed AUs & $52.3\%$ & $51.4\%$ & $51.8\%$ & $56.1\%$ & $51.1\%$ & $53.6\%$ \\ \hline
\end{tabular}
\label{tab:accuracy}
\end{table*}

%